\documentstyle[prb,preprint,aps]{revtex}
\begin{document}
\tighten
\preprint{cond-mat/9305007, submitted to \prb}
\title{
Electronic Structure and Bonding in Epitaxially Stabilized
Cubic Iron Silicides
}
\author{Kurt A.~M\"ader\cite{KAM} and Hans~von~K\"anel}
\address{
Laboratorium f\"ur Festk\"orperphysik, Eidgen\"ossische Technische
Hochschule H\"onggerberg, CH-8093 Z\"urich, Switzerland
}
\author{and}
\author{Alfonso Baldereschi\cite{AB}}
\address{
Institut Romand de Recherche Num\'erique en Physique des Mat\'eriaux,
Ecublens, CH-1015 Lausanne,\\
and Institut de Physique Appliqu\'ee, Ecole Polytechnique F\'ed\'erale
Lausanne, CH-1015 Lausanne, Switzerland.
}

\date{Received: 1 February 1993}
\maketitle
\begin{abstract}
We present an {\it ab initio\/} full-potential linearized augmented
plane-wave (FLAPW) study of the structural and electronic properties
of the two bulk unstable compounds FeSi (CsCl structure) and FeSi$_2$
(CaF$_2$ structure) which have recently been grown by molecular beam
epitaxy on Si(111).  We obtain equilibrium bulk lattice constants of 2.72
\AA\ and 5.32 \AA\ for FeSi and FeSi$_2$, respectively.  The density
of states (DOS) of FeSi agrees well with experiment, and shows
metallic behavior.  In agreement with a previous calculation the DOS
of FeSi$_2$ shows a large density of $d$-states at the Fermi level,
explaining the instability of the bulk phase.  The electron charge
distributions reveal a small charge transfer from Si to Fe atomic
spheres in both compounds.  While in FeSi the Fe-Si bond is indeed
partially ionic, we show that in FeSi$_2$ the electron distribution
corresponds to a covalent charge accumulation in the Fe-Si bond
region.  The reversed order of $d$-bands in FeSi with respect to FeSi$_2$ is
understood in terms of crystal field splitting and Fe-Fe nearest
neighbor $dd$-interactions in the CsCl structure, and a strong Si
$p$/Fe $d$ bonding in the fluorite structure, respectively.
\end{abstract}

\narrowtext
\section{Introduction}

The current interest in transition-metal silicides stems from both
fundamental and technological issues.  For applications in
microelectronics and optoelectronics the main focus is on silicides
which can be grown epitaxially on Si such as, e.g., NiSi$_2$ and
CoSi$_2$ whose small lattice mismatch with Si allows for the synthesis
of high quality semiconductor/metal heterostructures~\cite{Kaenel92}.
These systems open the way for the study of intriguing fundamental
questions, such as Schottky barrier formation, interface structure,
growth kinetics, and stability of novel epitaxial phases. Whereas the
CaF$_2$ structure of both CoSi$_2$ and NiSi$_2$ are stable bulk phases
of these compounds, FeSi$_2$ does not crystallize in the fluorite
form. At this stoichiometry and below $\sim 1000$\mbox{$^\circ$C}\ its
stable bulk form is \cite{Pearson,Binary} $\beta$-FeSi$_2$, which is
semiconducting. This orthorhombic structure can be regarded as a
distorted fluorite structure, and its greater stability with respect
to the fluorite phase can be understood in terms of the electronic
structure of the latter: a high density of $d$ states ($d$-DOS) at the
Fermi level $E_F$ leads to a Jahn-Teller-like instability of fluorite
\cite{Christensen90}
FeSi$_2$. In $\beta$-FeSi$_2$ the $d$ levels are crystal field split
at $E_F$, and the opening of a gap
\cite{Christensen90} ($E_g \approx 0.8$ eV) leads to semiconducting
behavior.  At composition 1{:}1 in the Fe-Si phase diagram
\cite{Binary} we find the compound $\epsilon$-FeSi. This structure is
actually referred to as the ``FeSi'' structure, and its Pearson symbol
is $cP8$.

Under the growth conditions of molecular beam epitaxy (MBE) on Si(111)
the Fe-Si phase diagram is dramatically different
\cite{Kaenel92b,Onda92,Onda93}: at a Fe/Si composition ratio of 1{:}2
not only epitaxial $\beta$-FeSi$_2$ is found, but also the bulk
unstable fluorite phase, denoted $\gamma$-FeSi$_2$ in Ref.\
\onlinecite{Onda92}. At composition 1{:}1 besides $\epsilon$-FeSi the CsCl
structure of FeSi grows pseudomorphically on Si(111). Hence we deal
with two novel materials completely absent in the bulk phase diagram.
Before turning to the electronic and structural properties of these
two structures, let us briefly discuss the physical mechanisms
pertinent to the occurrence of novel epitaxial phases. (We shall use
the term ``epitaxial phase'' only for novel phases which do not exist
in bulk form, in order to emphasize their distinctiveness from stable
bulk phases which may also be present in epitaxial phase diagrams.)

We distinguish two classes of epitaxial phases: ({\it i\/}) epitaxially
stable phases, where the strain conditions imposed by the substrate
select a species with a lower epitaxial formation enthalpy than the
strained bulk phase at that particular substrate lattice constant
\cite{Wood89,Bruinsma86}; ({\it ii\/}) epitaxially metastable phases,
where kinetic barriers inhibit the transformation to a (epitaxially)
more stable phase.

For 1{:}1 FeSi, we argue that up to a critical film
thickness the CsCl structure of FeSi belongs to class ({\it i\/}), i.e.,
it is epitaxially stable on Si(111).  Experimentally
\cite{Onda93} it is found that low temperature MBE growth of FeSi on Si(111)
yields coherent films in the CsCl structure up to $\sim 100$ \AA,
above which the creation of misfit dislocations leads to partial
relaxation. Films thicker than 15 \AA\ transform into
relaxed $\epsilon$-FeSi upon annealing at $\sim 300$\mbox{$^\circ$C}. The
experimentally determined lattice mismatches with Si(111) are $+2$\%
for FeSi (CsCl) and $-6.4$\% for $\epsilon$-FeSi, respectively.  We
can rationalize these experimental findings as follows (see Fig.\
\ref{fig:epi}): Due to the large lattice mismatch, pseudomorphic
$\epsilon$-FeSi will have a very high strain energy, and a small
critical thickness $h_1$, whereas the much smaller mismatch between
FeSi (CsCl) and the Si substrate makes this material more stable at
small thickness\cite{Wood89}. Since $\epsilon$-FeSi is the stable bulk phase,
however, above a critical thickness $H_c$ where it is completely
relaxed, it is lower in energy than the CsCl structure at any
thickness. Therefore, at some thickness $h_2$ the two energy curves
cross, and the CsCl structure ceases to be the epitaxially stable
phase. From experiment we infer $h_2 \lesssim H_c \approx 15$ \AA. Due
to kinetic barriers (e.g., an energetically favorable interface
structure), the CsCl phase is observed in metastable form even above $h_2$,
and beyond its
critical thickness $h_3\approx 100$
\AA\ up to $\sim 1000$ \AA.

At composition ratio 1{:}2 the
situation is less clear.  Many epitaxial relations for
$\beta$-FeSi$_2$/Si(111) with different lattice mismatches are
reported \cite{Sirring92}, and a disordered Fe$_{1-x}\Box_x$Si
phase of CsCl symmetry with an iron vacancy concentration $x=0.5$
acts as a precursor to the ordered CaF$_2$ phase
\cite{Onda93,Sirring92}. We will show, however,
that the CaF$_2$ phase is even
better lattice matched to Si than the CsCl phase, and hence a similar
picture as sketched in Fig.\ \ref{fig:epi} may apply also in this case.

Whereas the existence of epitaxial fluorite FeSi$_2$ is not that surprising
in view of its relation to the bulk stable $\beta$-phase and the existence
of CoSi$_2$ and NiSi$_2$ in this crystal structure, the CsCl structure of
FeSi seems to be rather ``exotic''.
It is interesting to note that the only transition metal silicides with
composition $M$Si which occur in the CsCl structure are those with
\cite{Finnie62} $M$
= Ru, Os, and Rh.  Ruthenium and Rhodium are the transition metals
situated in the Pd row of the periodic table just below Fe and Co,
respectively.  These silicides normally order in the ``FeSi''
structure type, i.e., the structure of $\epsilon$-FeSi, and only under
certain growth conditions do they form the CsCl structure
\cite{Finnie62}.  For instance, in the case of RuSi it was observed
that the CsCl structure formed ``more easily in mixtures that had not
yet reached equilibrium'' \cite{Finnie62}.  We have argued above that
FeSi films are epitaxially metastable above $\sim15$ \AA, and only
grow under the out-of-equilibrium conditions of low temperature MBE.
These considerations suggest that although the CsCl structure is
extremely rare for mono-silicides, it is nevertheless most likely to
be formed by certain group VIII transition metals, among which we find
Fe and Co.  Indeed, in Villars' three-dimensional structural stability
diagram \cite{Villars83}, both compounds are found in regions of
coexistence of the MnP-FeSi-CsCl structure types, but almost on the
boundary with the region occupied by the CsCl structure.  Very
recently, epitaxial growth of CsCl-type CoSi on Si(111) has indeed
been accomplished by MBE \cite{Kaenel92c}.  The existence of CsCl-type
NiSi, however, seems less probable.

The great versatility of Fe to form a number of epitaxial silicides on
Si---among which are found metals (FeSi(CsCl),$\gamma$-FeSi$_2$),
semiconductors or semimetals ($\beta$-FeSi$_2$, $\epsilon$-FeSi), and
even magnetic compounds (Fe$_3$Si, possibly
$\gamma$-FeSi$_2$)---offers a wide spectrum of potential applications.
Theoretical studies of the electronic and structural properties of the
Fe silicides are therefore of great interest.  Semiconducting
$\beta$-FeSi$_2$ has been studied by the linear muffin-tin orbital
(LMTO) \cite{Christensen90}, and by the augmented spherical-wave
\cite{Eppenga90} methods.  In Ref.\ \onlinecite{Christensen90}, the
electronic structure of the fluorite phase $\gamma$-FeSi$_2$ is also
given.  The magnetic compound Fe$_3$Si ($DO_3$ structure) has been
studied theoretically by Kudrnovsk\'y {\it et al.}
\cite{Kudrnovsky91}  We have recently published a brief account of the
electronic band structure of FeSi in the CsCl structure, together with
experimental results~\cite{Kaenel92b}.

In this work we concentrate on the properties of the bulk unstable
``high symmetry'' compounds FeSi (CsCl structure) and
$\gamma$-FeSi$_2$ (CaF$_2$ structure).  We determine their lattice
constants in order to better understand their existence as epitaxial
phases on Si(111), and give a detailed account of their electronic
band structures.  We focus on the bonding between Fe and Si both in
FeSi and $\gamma$-FeSi$_2$ and compare to NiSi$_2$ and CoSi$_2$.
Special attention is given to the covalent versus ionic nature of the
Fe-Si bond, and the role of Si $p$ - Fe $d$ interaction.

The paper is organized as follows.  In Section~\ref{sec:Total} we
present the details of our {\it ab initio} calculations and the
resulting equilibrium structural parameters of the two hypothetical
bulk phases; the Fe-Si bond is studied in Sec.~\ref{sec:Bond}; in
Sec.~\ref{sec:Elec} the electronic band structures are discussed; and
finally the conclusions are given in Sec.~\ref{sec:Conc}.

\section{Self-Consistent Calculations}
\label{sec:Total}

FeSi is assumed in the CsCl structure (space group $Pm\bar3m$), i.e.,
in a simple cubic lattice with Fe at the origin and Si at the body
center.  Both types of atoms are coordinated with 8 nearest neighbors
($nn$) of the other species.  $\gamma$-FeSi$_2$ crystallizes in the
CaF$_2$ structure (space group $Fm\bar3m$), i.e., in a face-centered
cubic lattice with Fe at the origin, and two Si atoms at positions
$\pm(\frac{1}{4}\frac{1}{4}\frac{1}{4})$, respectively.  Each Si atom
is tetrahedrally coordinated with 4 metal atoms, and each metal atom
has 8 Si $nn$.

We have calculated the electronic energy eigenvalues, total energies,
and wave functions of the two crystals with the self-consistent
full-potential linearized augmented plane-wave (FLAPW) method
\cite{Jansen84}, neglecting spin-polarization.
The exchange-correlation energy is treated in the
local density approximation (LDA), using the Hedin-Lundqvist
prescription.  Wave functions are expanded in spherical harmonics with
$l \le 8$ ($l\le 4$ for total energy minimization) inside the atomic
spheres whose radii are 1.85 a.u.\ and 2.3 a.u.\ for Si and Fe,
respectively.  In the interstitial region a plane wave basis is used
with wave vectors up to 3.6 Ry (8.5 Ry)  for wave functions
(charge and potential).  Integration in {\boldmath $k$}-space is
approximated by summing over up to 48 {\boldmath $k$}-points in the
irreducible wedge of the corresponding Brillouin zone.  The energy
band structure and density of states (DOS) of the equilibrium
structures are evaluated using higher cutoffs and more {\boldmath
$k$}-points.

The lattice constants and bulk moduli obtained by total energy
calculations are summarized in Tab.\ \ref{tab:struct}, together with
the lattice constants of the bulk phases.  We have fitted
the total energies calculated at six different lattice constants to
both Murnaghan's equation of state
\cite{Murnaghan44}, and a third- or fourth-order polynomial.  The
predicted equilibrium lattice constant is independent of the fitting
procedure, whereas the fluctuations of the fitted bulk modulus provide
the theoretical error bars given in the Table.  The theoretical values
predict a virtually perfect lattice match of CsCl-type FeSi and Si.
Recent X-ray diffraction data \cite{Onda93} on thin epitaxial
FeSi films yield a lattice constant $a_{\rm FeSi}^{\rm exp} = 2.77$
\AA, which is 1.8 \% larger than our theoretical value, and a larger
lattice mismatch with Si, i.e., $2 a_{\rm FeSi}^{\rm exp} = 5.54$ \AA\
($a_{\rm Si}^{\rm exp} = 5.43$ \AA) which remains however rather
small.  The predicted lattice constant of $\gamma$-FeSi$_2$ is only 2
\% smaller than that of FeSi (Tab.\ \ref{tab:struct}), even though
there are only half as many iron atoms per mole as in FeSi.  Assuming
the same magnitue of error on our predicted lattice constant as for the
CsCl structure, we obtain a corrected value of 5.42 \AA,
which is almost perfectly
lattice matched with Si. This explains why the fluorite phase can compete
with the lattice mismatched bulk phase ($\beta$-FeSi$_2$) in the presence
of the Si substrate (see our discussion in the Introduction).
While the Fe-Si $nn$ distance differs by only 2 \% in the two
crystals, the Fe-Fe $nn$ distance is 2.72 \AA\ in FeSi, and 3.76 \AA\
in FeSi$_2$.  Note that this Fe-Fe spacing in FeSi is only 10
\% greater than the $nn$ distance, and 5 \% smaller than
the next-nearest neighbor distance along the cube edge in {\it bcc}
iron.

Given the near lattice matching of both FeSi (CsCl) and FeSi$_2$
(CaF$_2$) with Si, we shall describe the electronic properties of the
{\em unstrained} hypothetical equilibrium phases, rather than the
rhombohedrally distorted ones observed on Si(111). The strain induced
changes on the electronic band structure are expected to be very
small.

\section{The F\lowercase{e}-S\lowercase{i} Bond}
\label{sec:Bond}

The nature of the chemical bond in transition metal silicides has been
the subject of extensive theoretical and experimental studies
\cite{Chabal82,Tersoff83,Bisi81,Lambrecht87,Weaver84}.  In general, the bonding
in the silicides is understood in terms of metal $d$ - silicon $p$
hybridization.  There has been a controversy about the bonding in
NiSi$_2$.  Tersoff and Hamann \cite{Tersoff83} proposed that the
bonding in NiSi$_2$ and CoSi$_2$ is similar to that in Si, an idea
first put forward by Chabal {\it et al.\/}
\cite{Chabal82}  Following their arguments, the tetrahedrally coordinated Si
atoms build $sp^3$ hybrids, which in the silicides form directed
covalent bonding states with Fe $d$ orbitals.  In a later study,
Lambrecht {\it et al.\/} \cite{Lambrecht87} came to a similar
conclusion, but attributed more importance to the charge transfer from
Si to Ni which contributes an electrostatic term to the cohesive
energy.  It is well known, however, that the charge transfer is not
uniquely determined, since it depends on the choice of the ``atomic
spheres'' where the charge is measured.  With this in mind, we compare
in Tab.\ \ref{tab:charge} the valence charge contained in the atomic
spheres used in our calculations.  Besides the self-consistent density
of the two crystals we have calculated the charge density of the
isolated atoms as well as the lattice superpositions of the atomic
densities.  In FeSi$_2$, the 8 Si {\it nn} of Fe contribute more
charge to the Fe spheres due to the smaller Fe-Si distance.  On the
other hand, the six Fe {\it nn} of Fe in FeSi are much closer than the
12 Fe {\it nn} of Fe in FeSi$_2$, therefore superposition of atomic
densities gives a Fe charge 0.08 electrons {\em larger} in FeSi than
in FeSi$_2$.  As to the charge in the Si spheres it is obvious that
twice the number of Fe {\it nn} of Si in FeSi yields a larger overlap
than the corresponding one in FeSi$_2$.  The remaining free-atom
charge is distributed in the interstitial region.  When comparing the
free-atom superpositions with the self-consistent densities, we
observe an additional {\em net} charge transfer $\Delta Q_{\rm Fe}$ of
0.18 electrons into the Fe sphere in FeSi, and one of 0.35 electrons
into the Fe sphere in FeSi$_2$.  Note that now the Fe sphere in
FeSi$_2$ contains more charge than the one in FeSi.  It would be
misleading, however, to draw conclusions on the ionicity from these
numbers alone, due to the ambiguity pointed out above.  We will show
that the excess charge in the FeSi$_2$ Fe sphere with respect to the
FeSi one is due to a piling up of covalent charge in the Fe-Si bond
region of FeSi$_2$.

To this end we
analyze the charge {\it densities\/} instead of the
integrated charge in the spheres.  In Fig.\ \ref{fig:rho} we display
contour plots of the difference between the self-consistent charge
density and the superposition of atomic densities in the (110) plane
of FeSi$_2$ (a) and FeSi (b).  Besides the obvious difference due to
the presence of an additional Fe atom in FeSi, we notice a weak piling
up of charge midway between Fe and Si in the fluorite structure.  This
covalent charge accumulation is almost entirely contained in the Fe
sphere, and is  responsible for the larger charge transfer in
FeSi$_2$ discussed above.  In FeSi, on the other hand, the charge
re-distribution is unimportant in the interstitial and bond region,
and we interpret the charge transfer of 0.18 electrons from the Si to
the Fe spheres as a real ionic character of the bond.  This conclusion
will be reinforced by the crystal field arguments presented in the
next section.

\section{Electronic Structure}
\label{sec:Elec}

The energy dispersion relations of $\gamma$-FeSi$_2$ and CsCl-type
FeSi can be found in Refs.\ \onlinecite{Christensen90} and
\onlinecite{Kaenel92b}, respectively, and are not reproduced here.  In
Fig.\ \ref{fig:DOS} we show the DOS of the two materials
instead\cite{LDAnote}.  In Tab.\ \ref{tab:DOS} we provide the
angular-momentum projected densities at $E_F$ and integrated up to
$E_F$, respectively.  Our DOS for FeSi$_2$ is in perfect agreement
with the LMTO one by Christensen
\cite{Christensen90}.  The two main peaks in the valence spectrum
originate from the metal $d$/Si $p$ bonding states at $\sim -4$ eV
($\Gamma_{25'}^v$ states at the zone center) and from the non-bonding
metal $d$-band manifold centered at $\sim -1.5$ eV ($\Gamma_{12}^v$
states at the zone center), respectively.  LDA energy eigenvalues at
high symmetry points are listed in Tab.\ \ref{tab:FeSi2}, together
with the angular-momentum decomposition of the wave functions inside
the atomic spheres.  The $pd$ hybridization broadens the metal $d$
bands considerably as can be seen from the DOS and the orbital
characters of the eigenfunctions.  Given the strong similarity between
the present Si $s$ and $p$ partial DOS with those of CoSi$_2$
\cite{Tersoff83}, we conclude that the binding mechanism in FeSi$_2$
is basically the same as in the former compound.  It has been
shown by Christensen that the rigid band assumption is nearly
fulfilled for the three compounds FeSi$_2$, CoSi$_2$ and NiSi$_2$
(see Fig.\ 1 in Ref.\
\onlinecite{Christensen90}), and therefore the number of valence electrons
pins the position of the Fermi level in an otherwise similar DOS.  At
the Fermi level there is a rather strong peak in the $d$-DOS of
FeSi$_2$ (28.5 Ry$^{-1}$, cf.\ Tab.\ \ref{tab:DOS}),
with an appreciable admixture of Si $p$-states.  The location in
{\boldmath $k$}-space of these states is shown in Ref.\
\onlinecite{Christensen90} to be on the hexagonal Brillouin-zone face.  We
have previously estimated \cite{Onda92} that the Stoner factor is
considerably larger than unity and explained bulk instability of the
fluorite phase in terms of the Stoner model.  Indeed, a spin-polarized
calculation by Christensen predicts a ferromagnetic moment of 0.3
$\mu_B$ per formula unit \cite{Christensen90}.  Experimental
confirmation of ferromagnetism in $\gamma$-FeSi$_2$ is still lacking,
since the grown films are very thin ($\le 15$ \AA) and magnetic
measurements are difficult to perform.  Above $E_F$ we observe a Si
$sp$ antibonding peak at 3 eV, which seems to be even more pronounced
than in CoSi$_2$ \cite{Tersoff83}, and which is a typical fingerprint
of $sp^3$ hybridization.  Together with the discussion of the charge
transfer in Sec.\
\ref{sec:Bond} the picture of covalent bonding in fluorite FeSi$_2$ is
thus confirmed, similarly to the case of CoSi$_2$ and NiSi$_2$, and in
agreement with the interpretation of Tersoff and Hamann \cite{Tersoff83} for
the latter two compounds.
In FeSi$_2$, however, the covalent contribution to the free energy
is too small to stabilize the fluorite phase in bulk form. The kinetics and
epitaxial constraints of iron silicide grown by MBE on Si(111), however,
lead to the formation of fluorite FeSi$_2$ as an intermediate step of
the disordered Fe$_{0.5}\Box_{0.5}$Si to $\beta$-FeSi$_2$
transition \cite{Kaenel92b,Onda93} at $\sim 500$\mbox{$^\circ$C}.

Let us now turn to FeSi (right panel in Fig.\
\ref{fig:DOS}, and Tab.\ \ref{tab:FeSi}).
Again the total DOS is dominated by the metal $d$ bands, giving rise
to a strong 2 eV broad peak centered at $\sim -2$ eV.  The Fermi level
is located at the upper end of a region of low DOS, at the onset of a
less pronounced, sharp peak of Fe $d$ states.  The valence DOS agrees
very well with the ultraviolet photoemission spectrum of Ref.\
\onlinecite{Kaenel92b}, provided that the theoretical curve is shifted
by $\sim1$ eV towards $E_F$.  This final-state effect is discussed in
more detail in Ref.\
\onlinecite{Kaenel92b}.  A very similar DOS is also predicted for CsCl-type
RuSi by a LMTO Green's function calculation of Ivanovskii
\cite{Ivan90}.
There it is also shown that RhSi has a virtually identical DOS,
except for the position of $E_F$ which now falls at higher energy in
the sharp $d$ peak.  The same behavior can therefore be anticipated
when replacing Fe by Co in FeSi, indicating that the hypothetical CsCl
phase of CoSi is even less stable than FeSi, and likely to be
magnetic.  Below $E_F$ the Si $s$ and $p$ projected DOS is rather
similar to that in FeSi$_2$.  This does not imply the formation of
$sp^3$ hybrids, however, since Si is now 8-fold coordinated, instead
of 4-fold as in the fluorite structure.  We attribute the relatively
large width of the Si $s$ band to bonding with Fe $s$ and $p$ states
(see Tab.\ \ref{tab:FeSi}).

An important difference between the band structures of the two
materials lies in the order of the Fe $d$ bands at $\Gamma$: in FeSi
$\Gamma_{25'}$ lies above $\Gamma_{12}$, whereas in FeSi$_2$
$\Gamma_{25'}$ lies below $\Gamma_{12}$. An inverted crystal field
splitting was previously noted in band calculations of chalcopyrites
\cite{Jaffe83} (CuInSe$_2$) and in II-VI compounds \cite{Wei88}. The
results were explained in terms of symmetry-enforced $pd$ couplings. A
similar interpretation is pertinent here: In FeSi the Si $p$-bands
$(x,y,z)$ transform like $\Gamma_{15}$, and thus do not mix with the
$d$-band triplet $(xy,yz,zx)$ of Fe, which transforms like
$\Gamma_{25'}$. Hence, the crystal field splitting corresponds to
$\Gamma_{12}$ below $\Gamma_{25'}$ like predicted by the point-ion
model.  This mixing is allowed, however, in the fluorite structure,
where the $p$-orbitals of the two Si atoms in the unit cell originate
both $\Gamma_{15}$ and $\Gamma_{25'}$ states.  The resulting $pd$
bonding lowers the $\Gamma_{25'}$ level relative to the non-bonding
$\Gamma_{12}$ level, thus reversing the sign with respect to that
originating from the crystal field splitting alone.  This effect is
illustrated in Fig.\
\ref{fig:pd}, where the energy levels of the Fe $d$ and Si $p$ states
are shown for FeSi$_2$ in the CaF$_2$ structure, (a) without and (b)
with taking the $pd$ interaction into account. The $pd$ repulsion
found in certain II-VI zincblende semiconductors has a similar origin
\cite{Wei88}: the cation $d$ bands of $\Gamma_{15}$ symmetry mix with
the anion $p$ bands and are pushed below the $\Gamma_{12}$ $d$ states,
opposed to crystal field considerations alone. Note, however, that the
$pd$ repulsion in ZnTe, CdTe, and HgTe, e.g., is much weaker than in
the present case of FeSi$_2$ and in the case of the chalcopyrites,
since ({\it i\/}) the cation $d$ - anion $p$ energy separation is larger
for these atom pairs than for Fe and Si, ({\it ii\/}) the cation-anion
distance is considerably larger than the Fe-Si one in FeSi$_2$.
Therefore, the $\Gamma_{15} - \Gamma_{12}$ splittings are much smaller
in magnitude in these II-VI compounds \cite{Wei88} than the
$\Gamma_{25'} -
\Gamma_{12}$ splitting in FeSi$_2$, which we predict to be $-2.55$ eV.

In FeSi, on the other hand, $E(\Gamma_{25'}) - E(\Gamma_{12}) = +3.24$ eV,
which is an unusually large crystal field splitting.
We argue that it has two major contributions, an ionic one and one due to
repulsion
between  Fe $d$ states of neighboring Fe atoms.
Placing positive point charges
$+ze$ on the Si atom sites and negative charges $-ze$ on the Fe sites
will lower the electrostatic binding energy of the $\Gamma_{12}$
states relative to the $\Gamma_{25'}$ ones.  We estimate the ionic
contribution $\Delta E_i$ to the splitting by assuming analytic Slater
orbitals for the Fe $3d$ wave functions, and by performing the lattice
sum involved in the fourth-order non-spherical portion of the crystal
potential as described by Callaway \cite{Callaway91} for cubic
crystals.  We obtain $\Delta E_i \approx +1.14 z$ eV for
Fe$^{-z}$Si$^{+z}$.  Clearly, this  crude point-charge ionic model is
not sufficient to quantitatively explain the level splitting, but the
important points are that the sign agrees with the calculated one and
that ionicity contributes to the splitting.  A similar calculation for
the fluorite structure (placing the charge $-2ze$ on the Fe sites)
yields a $d$ level splitting with {\it equal\/} sign as in FeSi, but 40 \%
smaller.  This is not surprising, since the nearest neighbor
configuration of the Fe atoms is the same for both structures, and the
contribution from the outer shells decreases rapidly
\cite{Callaway91}.  Hence, neglecting $p$-$d$ bonding in the fluorite
structure the $\Gamma_{12}$ states would be {\it lower\/} in energy
than the $\Gamma_{25'}$ ones (Fig.\ \ref{fig:pd}(a)), very much like in
the CsCl structure.  This shows once more that ionicity is not
important in FeSi$_2$.

The other effect which contributes to the large splitting in FeSi is
the Fe-Fe interaction.  We noted in Sec.\ \ref{sec:Total} that the
Fe-Fe {\it nn\/} distance in FeSi is rather close to the equilibrium
bond length in {\it bcc\/} iron.  According to Harrison
\cite{Harrison80} the $dd$ interatomic matrix elements decrease with
the 5th power of the distance, and should therefore be about 60 \% as
large as in {\it bcc\/} iron.  Recent tight-binding fits to the
present band structures show indeed that the Fe-Fe $dd$ interaction
cannot be neglected in FeSi, whereas it is less important in
\cite{Miglio92} FeSi$_2$.  In FeSi the Fe atoms occupy simple cubic lattice
points, and hence are octahedrally coordinated to their nearest Fe
neighbors.  The $\Gamma_{12}$ orbitals form $\sigma$-bonds, whereas
the $\Gamma_{25'}$ orbitals are $\pi$-bonded.  Using Harrison's
universal tight-binding parameters \cite{Harrison80} a splitting of
$\sim 2.2$ eV is obtained.  Again, this is not sufficient by itself to
explain the splitting obtained in the self-consistent calculation.
Considering in addition the ionic contribution discussed above, however,
(calculated with a better approximation than the point-charge model)
one should recover the $\Gamma_{25'}$-$\Gamma_{12}$ splitting of the FLAPW
calculation.

Inspection of Tab.\ \ref{tab:FeSi} and of Fig.\
\ref{fig:DOS} (right panel) shows that Si $p$/Fe $d$
hybridization is present also in FeSi, although to a smaller extent
than in FeSi$_2$, at {\boldmath $k$}-points other than $\Gamma$.  At
the $R$-point, e.g., the {\boldmath $k$}-dependent phase factors
associated with the atomic orbitals in a LCAO scheme cause the Si
$(x,y,z)$ - like and the Fe $(xy,yz,zx)$ - like wave functions to
belong to the same irreducible representation $R_{15}$, as they do
under the $T_d$ point group.  Also along the $\Lambda$-line $pd$
hybridiztion is allowed by symmetry, and leads to anticrossing (see
Fig.\ 4 in Ref.\ \onlinecite{Kaenel92b}) of the twofold degenerate
$\Lambda_3$ bands originating from both $\Gamma_{12}$ and
$\Gamma_{25'}$ (retaining only $dd$ matrix elements, these bands would
cross along $\Lambda$).  Due to $pd$-bonding the $R_{15}$ state is
pushed below $\Gamma_{12}$, and contributes to the peak in the DOS at
$\sim -6$ eV (see Fig.\ \ref{fig:DOS}).  The upper $\Lambda_3$ band
ends in a ``$dd$-antibonding'' $R_{12}^c$ state close to the Fermi
level, with 99 \% $d$-character in the Fe atomic spheres.  Due to the
strong $dd$ interaction we refrain from using the term ``non-bonding
$d$'' for these states.

\section{Conclusions}
\label{sec:Conc}

We have presented the first theoretical study of the electronic and
structural properties of the bulk unstable cubic iron silicide
FeSi(CsCl), and have compared them with those of
$\gamma$-FeSi$_2$(CaF$_2$). Both structures exist epitaxially on
Si(111) due to a lattice mismatch with Si wich is smaller than that of
the bulk stable phases $\epsilon$-FeSi and $\beta$-FeSi$_2$,
respectively. In addition, kinetic barriers are believed to exist which
account for the observed metastability of FeSi (CsCl) films at thicknesses
exceeding its epitaxial stability range (Fig.\ \ref{fig:epi}).

We have established the close resemblance of the
bonding configuration of FeSi$_2$ with CoSi$_2$ and NiSi$_2$, i.e., Si
$sp^3$ hybridization, strong metal $d$/Si $p$ bonding, the presence of
directed covalent Fe-Si bonds, and negligible ionicity.  The giant
density of $d$ states at the Fermi level leads to a Jahn-Teller
instability of the bulk compound, explaining its absence in the bulk
phase diagram.

CsCl-type FeSi, on the other hand, is characterized by a small charge
transfer from Si to Fe, strong Fe-Fe $dd$ bonding, and to some extent
also by Si $p$/metal $d$ interaction.  We have shown that both the
charge distribution and the electronic band structure of the two
materials are consistent with the bonding mechanisms presented in this
work.  An empirical tight-binding study of the Fe silicides is under
way \cite{Miglio92}, and will address the lattice dynamics and its
implications on the $\gamma\to\beta$ phase transition in
epitaxial FeSi$_2$/Si(111).

Comparison with other known monosilicides in the CsCl structure has
allowed us to speculate on the possible existence of epitaxial
CoSi(CsCl) on Si(111).  Subsequent experiments have confirmed the
epitaxial stability of this phase.  A rigid-band model implies a
rather high $d$-DOS at the Fermi level, which may cause the compound
to order magnetically.

The disordered CsCl phase Fe$_{1-x}\Box_x$Si is observed over a
wide range of Fe vacancy concentrations $0< x\le 0.5$.  From the
present study of the ordered end compounds FeSi and FeSi$_2$ we
presume that the ionic, or Madelung contribution to the cohesive
energy (present in the CsCl, but not in the CaF$_2$ phase) stabilizes
the epitaxial disordered Fe$_{0.5}\Box_{0.5}$Si CsCl structure
relative to the ordered CaF$_2$ structure. In the latter the Fe atoms
are located on alternate
\{111\} planes, and the Si atoms are tetrahedrally coordinated with the
Fe atoms.  The role of the $d$ electrons, however, cannot be obtained
from the properties of the end compounds alone.  A rigid-band model
would predict the Fermi level to fall in a region of high $d$-DOS for
increasing $x$, which is hardly reconciled with the experimentally
observed epitaxial stability and absence of magnetic order in the
defect CsCl phase.  Further theoretical studies using, e.g., the
coherent phase approximation, are needed for a complete understanding
of the electronic properties of Fe$_{1-x}\Box_x$Si as a function
of $x$.  We hope that the present study, which adds to the general
understanding of transition metal silicides, will stimulate further
theoretical and experimental work on the fascinating iron silicide
system.

\acknowledgments
The authors are indebted to M.\ Posternak for the permission to use
his FLAPW code, to F.\ Hulliger for drawing their attention to the
existence of CsCl-type mono-silicides other than FeSi, and to A.\
Zunger and S.\ Froyen for a critical reading of the manuscript and
helpful comments.  Financial support by the Board of the Swiss Federal
Institutes of Technology, and in part by the ``Stiftung
Entwicklungsfonds Seltene Metalle'' is gratefully acknowledged.  The
calculations have been performed on the Cray Y-MP at ETH Z\"urich.


\narrowtext
\begin{figure}
\caption{Schematic diagram of eptiaxial energy versus film thickness for
FeSi on Si(111). The CsCl structure is more closely matched to the Si
substrate than the bulk stable $\epsilon$-phase, and hence is
epitaxially stable at small film thickness. Above its critical
thickness $H_c$ $\epsilon$-FeSi is the most stable phase, and the
occurrence of the CsCl phase in this range can be explained by the
existence of a kinetic barrier.  Upon annealing at $\sim
300$\mbox{$^\circ$C}\ the CsCl phase transforms irreversibly into the
bulk $\epsilon$-phase.
\label{fig:epi}}
\end{figure}

\widetext
\begin{figure}
\caption
{Contour plot of the  difference between the
self-consistent charge density and the superposition of atomic
densities in the (110) plane of CaF$_2$-type FeSi$_2$ (a) and
CsCl-type FeSi (b). Areas of negative values are shaded, contours are
labeled in units of $10^{-2}$ electrons/(bohr)$^3$, and the level
spacing is 0.02 in these units.  $\Delta Q_\alpha$ is the difference
between the self-consistent valence charge and the superposition of the
atomic valence charges in the  sphere centered on atom $\alpha$.
The Fe-Si bonds are indicated with solid lines.
\label{fig:rho}}
\end{figure}

\widetext
\begin{figure}
\caption{DOS of CaF$_2$-type FeSi$_2$ (left panel) and CsCl-type FeSi (right
panel).  (a) and (c) show the total (solid line) and partial Fe $d$
(dashed-dotted line) DOS in eV$^{-1}$ per unit cell, (b) and (d) show
the partial Si $p$ (solid line) and Si $s$ (dashed line) DOS. The
partial DOS are projected on {\it one} atomic sphere in both
structures. The energy is measured relative to the Fermi level
(indicated by a solid vertical line).
\label{fig:DOS}}
\end{figure}

\narrowtext
\begin{figure}
\caption{Energy levels of Fe $d$ and Si $p$ bands at $\Gamma$ in the CaF$_2$
structure of FeSi$_2$, (a) neglecting, (b) including  Fe $d$/Si $p$
interaction. Note that $pd$ repulsion pushes the $\Gamma_{25'}$ states below
the $\Gamma_{12}$ states, whereas crystal field splitting alone (a) predicts
the $\Gamma_{12}$ states to be lowest.
\label{fig:pd}
}
\end{figure}


\mediumtext
\begin{table}
\caption{Structural definition of the stable bulk phases \protect\cite{Pearson}
of FeSi and FeSi$_2$,
as well as of the bulk unstable phases which exist epitaxially on Si(111).
For the latter the theoretical predictions of the present work are compared
to experimental values, where available, and to other theoretical results.
The experimental lattice constant of Si is 5.43 \protect\AA.}
\begin{tabular}{llddc}
Compound & Space group & \multicolumn{2}{c}{Lattice parameters (\AA)} &
  Bulk modulus (Mbar) \\
(Pearson Symbol) & & \multicolumn{1}{c}{(exp.)} & \multicolumn{1}{c}{(theo.)}
 &              \multicolumn{1}{c}{(theo.)} \\
\hline
Bulk stable: \\
$\epsilon-$FeSi & $P2_13$ & $a=$ 4.46\\
($cP8$) \\
$\beta-$FeSi$_2$ & $Cmca$ & $a=$ 9.86\\
($oC48$) &         & $b=$ 7.79\\
                &         & $c=$ 7.83\\
\hline
Bulk unstable: \\
FeSi (CsCl)	& $Pm\bar3m$ & $2a=$ 5.54\tablenotemark[1]    &$2a$ = 5.44
                                       &2.70 $\pm$0.07  \\
($cP2$)\\
Fe$_{1-x}\Box_x$Si & $Pm\bar3m$ & $2a=$ 5.40\tablenotemark[1]    &  &  \\
(CsCl, $x\le0.5$)			 &	      & ($x=$ 0.5)\\
$\gamma-$FeSi$_2$ (CaF$_2$) & $Fm\bar3m$ &   & $a=$ 5.32 &1.90 $\pm$0.05 \\
($cF12$) &                    &   & $a=$ 5.39\tablenotemark[2] &
       					2.06\tablenotemark[2]  \\
\end{tabular}
\label{tab:struct}
\tablenotetext[1]{Reference \protect\onlinecite{Onda93}.}
\tablenotetext[2]{Reference \protect\onlinecite{Christensen90}.}
\end{table}

\mediumtext
\begin{table}
\caption{Valence charge contained in the atomic spheres of radius $R$
and in the interstitial volume of CsCl-type FeSi and
CaF$_2$-type FeSi$_2$. The total valence charge of Fe is 8, and
that of Si is 4 electrons.
}
\begin{tabular}{llcccc}
        & Charge & Fe & Si & Interstitial & Total \\
	&	 & $R=2.3$ a.u. & $R=1.85$ a.u. & & \\
\hline
Fe, Si  & isolated atoms &\dec 6.24 &\dec 1.25 & & \\
FeSi	& free-atom superposition&\dec 7.30 &\dec 1.66 &\dec 3.04 &\dec 12.0 \\
	& self-consistent &\dec 7.48 &\dec 1.48 &\dec 3.04 &\dec 12.0 \\
FeSi$_2$& free-atom superposition&\dec 7.22 &\dec 1.55 &\dec 5.68 &\dec 16.0 \\
	& self-consistent&\dec 7.57 &\dec 1.42 &\dec 5.59 &\dec 16.0 \\
\end{tabular}
\label{tab:charge}
\end{table}
\bigskip

\mediumtext
\begin{table}
\setdec 0.000
\caption{Partial number of states (NOS) below the Fermi energy $E_F$, and
density of states at $E_F$ (DOS in Ry$^{-1}$)
normalized to one atomic sphere (cf.\ Tab.\ \protect\ref{tab:charge})
in CsCl-type FeSi and CaF$_2$-type FeSi$_2$.}
\begin{tabular}{llcccccc}
     &    & Fe $s$ & Fe $p$ & Fe $d$ & Si $s$ & Si $p$ & Si $d$ \\
\hline
FeSi & DOS&\dec 0.001 &\dec 0.018 &\dec 1.678 &\dec 0.051 &\dec 0.020 &\dec
0.004 \\
     & NOS&\dec 0.385 &\dec 0.530 &\dec 6.528 &\dec 0.599 &\dec 0.720 &\dec
0.143 \\
FeSi$_2$ &DOS&\dec 0.317 &\dec 3.740 &\dec 28.478 &\dec 0.073 &\dec 2.115
&\dec 0.449 \\
     & NOS&\dec 0.466 &\dec 0.540 &\dec 6.544 &\dec 0.607 &\dec 0.706 &\dec
0.099 \\
\end{tabular}
\label{tab:DOS}
\end{table}

\mediumtext
\begin{table}
\setdec -10.00
\caption{FLAPW energy eigenvalues at high symmetry points of the
face centered cubic
Brillouin zone for FeSi$_2$ in the CaF$_2$ structure.
Energies are measured relative
to the Fermi level. The probability of the orbital character of the
wavefuntions inside the Fe and Si atomic
spheres is given in percent. In the last column we list
the probability in the interstitial region.}
\begin{tabular}{lcccccccc}
State & Energy & \multicolumn{3}{c}{Fe} & \multicolumn{3}{c}{Si} & int. \\
      & (eV)   & s & p & d & s & p & d & \\
\hline
$\Gamma_1^v$    &\dec -13.07 & 14 & & & 30 & & & 56 \\
$\Gamma_{25'}^v$ &\dec -3.95 & & & 50 & & 17 & & 32 \\
$\Gamma_{12}^v$ &\dec -1.40  & & & 80 & & &  3 & 17 \\
$\Gamma_{2'}^c$ &\dec  0.87  & & & & 67 & & &    29 \\
$\Gamma_{15}^c$ &\dec  3.85  & & 28 & & & 2 & 13 & 56 \\
$X_{1'}^v$	&\dec -8.93 & & 10 & & 36 & &1 & 52 \\
$X_{1}^v$	&\dec -7.09  & 15 & & 8 & & 19 & & 58 \\
$X_{3}^v$	&\dec -6.43  & & & 28 & 33& &2 & 37 \\
$X_{5}^v$	&\dec -3.49  & &13 & & & 27 & & 59 \\
$X_{1}^v$	&\dec -1.44  &14 & & 52 & &1 &5 & 28 \\
$X_{5}^v$	&\dec -1.43  & & & 86 & & 4 &1 & 9 \\
$X_{2}^v$	&\dec -0.67  & & & 91 & & & & 8 \\
$X_{1}^c$	&\dec  2.60  & & & 74 & 16 & & &10 \\
$L_{1}^v$	&\dec -9.73  &13 & & 5& 26 & 5 & & 51 \\
$L_{2'}^v$	&\dec -9.16  & &10 & & 26 &6 & & 58 \\
$L_{3}^v$	&\dec -3.41  & & & 53 &  &13 &1 & 32 \\
$L_{1}^v$	&\dec -1.70  &9 & &63 &2 &5 & 2 & 17 \\
$L_{3}^v$	&\dec -1.04  & & & 87 & & & 1 & 11 \\
$L_{3'}^v$	&\dec -0.24  & & 16 & & & 26 & 4 & 53 \\
$L_{2'}^c$	&\dec  2.74  & & & & 23 & 11 & 3 & 59 \\
\end{tabular}
\label{tab:FeSi2}
\end{table}

\mediumtext
\begin{table}
\setdec -10.00
\caption{FLAPW energy eigenvalues at high symmetry points of the simple cubic
Brillouin zone for FeSi in the CsCl structure. Energies are measured relative
to the Fermi level. The probability of the orbital character of the
wavefuntions inside the Fe and Si atomic
spheres is given in percent. In the last column we list
the probability in the interstitial region.}
\begin{tabular}{lcccccccc}
State & Energy & \multicolumn{3}{c}{Fe} & \multicolumn{3}{c}{Si} & int. \\
      & (eV)   & s & p & d & s & p & d & \\
\hline
$\Gamma_1^v$    &\dec -14.60 & 24 & & & 27 & & & 48 \\
$\Gamma_{12}^v$ &\dec -4.12  & & & 76 & & &  3 & 20 \\
$\Gamma_{25'}^v$&\dec -0.88  & & & 95 & & &  1 &  3 \\
$\Gamma_{15}^c$ &\dec 3.38   & & 17 & & & 37 & & 40 \\
$R_{15}^v$	&\dec -6.30  & & & 50 & & 16 & & 34 \\
$R_{1'}^v$	&\dec -1.27  & & &    & 64 & & & 30 \\
$R_{12}^c$	&\dec 0.07   & & & 99 &    & & &  1 \\
$R_{15}^c$	&\dec 1.67   & & 39 & & & & 12 & 48 \\
$X_{2'}^v$	&\dec -10.47 & & 17 & & 34 & & & 48 \\
$X_{1}^v$	&\dec -9.04  & 26 & & 10 & & 17 & & 46 \\
$X_{3}^v$	&\dec -3.15  & & & 87 & & & &	 12 \\
$X_{5}^v$	&\dec -2.53  & & & 81 & &  6 & & 11 \\
$X_{4}^v$	&\dec -0.69  & & & 97 & &    & &  2 \\
$X_{3}^c$	&\dec 0.34   & 4 & & 89 & & 2 & & 4 \\
\end{tabular}
\label{tab:FeSi}
\end{table}


\begin{references}
\bibitem{LDAnote} Note that the DOS presented here is based on the LDA.
Self-energy corrections are known to shift energy eigenvalues below $E_F$
to lower energies in proportion to the degree of their spatial localization.
Thus, $d$ states will move more than $sp$ states. These corrections should not
affect our conclusions, however.
\end{references}
\end{document}